\documentclass[aps,prb,twocolumn,superscriptaddress,showpacs,floatfix]{revtex4}
\usepackage{graphicx}
\usepackage{amsmath}
\usepackage{amssymb}
\usepackage{url}

\graphicspath{{.}{../}{./eps/}}

\begin{document}
\title{Bound state formation and nature of the excitonic 
insulator phase in the extended Falicov-Kimball model}
\author{D. Ihle}\affiliation{Institut f\"ur Theoretische Physik, 
Universit\"at Leipzig, 04109 Leipzig, Germany}
\author{M. Pfafferott}\affiliation{Institut f\"ur Physik, 
Ernst-Moritz-Arndt Universit\"at
  Greifswald, 17487 Greifswald, Germany}
\author{E. Burovski}
\affiliation{\mbox{Laboratoire de Physique Th{\'e}orique et
Mod{\`e}les Statistiques, Universit{\'e} Paris-Sud, 91405 Orsay
Cedex, France}}
\author{F. X. Bronold}\affiliation{Institut f\"ur Physik, Ernst-Moritz-Arndt Universit\"at
  Greifswald, 17487 Greifswald, Germany}
\author{H. Fehske}
\affiliation{Institut f\"ur Physik, Ernst-Moritz-Arndt Universit\"at
  Greifswald, 17487 Greifswald, Germany}\date{\today}
\begin{abstract}
Motivated by the possibility of pressure-induced exciton condensation 
in intermediate-valence Tm[Se,Te] compounds we study the Falicov-Kimball 
model extended by a finite f-hole valence bandwidth. 
Calculating the Frenkel-type exciton propagator we obtain excitonic
bound states above a characteristic value of the local 
interband Coulomb attraction. Depending on the system parameters 
coherence between c- and f-states may be established  at low temperatures,
leading to an excitonic insulator phase. 
We find strong evidence that the excitonic insulator typifies either a 
BCS condensate of electron-hole pairs (weak-coupling regime)
or a Bose-Einstein condensate (BEC) of preformed excitons 
(strong-coupling regime), which points towards a BCS-BEC 
transition scenario as Coulomb correlations increase. 
   
\end{abstract}

\pacs{71.28.+d, 71.35.-y, 71.35.Lk, 71.30.+h, 71.28.+d. 71.27.+a}
\maketitle

That excitons in solids might condense into a macroscopic phase-coherent  
quantum state---the excitonic insulator---was theoretically proposed 
about more than four decades ago~\cite{Kno63}, for a recent review
see Ref.~\onlinecite{LEKMSS04}. The experimental 
confirmation has proved challenging, because excitonic quasiparticles 
are not the ground state but bound electron-hole excitations that 
tend to decay on a very short timescale. Thus a large number of excitons 
has to be created, e.g. by optical pumping, with sufficiently long lifetimes 
as a steady-state precondition for the Bose-Einstein condensate
(BEC) realizing process. 

The obstacles to produce a BEC out of the far-off-equilibrium 
situation caused 
by optical excitation might be circumvented by pressure-induced generation
of excitons. Hints that pressure-sensitive, narrow-gap semiconducting 
materials, such as intermediate-valent $\rm TmSe_{0.45}Te_{0.55}$, 
might host an excitonic BEC in solids came from a series of 
electric and thermal transport measurements.~\cite{NW90}
Fine-tuning the excitonic level, by applying pressure, 
to the level of electrons   in the narrow 
4f-valence band, excitons can form near the
semiconductor semimetal transition in thermodynamical equilibrium and might
give rise to collective excitonic phases. A phase diagram has been deduced 
out of the resistivity, thermal diffusity and heat conductivity data, 
which contains, below 20 K and in the pressure range between 5 and 11 kbar, 
a superfluid Bose condensed state.~\cite{WBM04}

The experimental claims for excitonic condensation in 
$\rm TmSe_{0.45}Te_{0.55}$ have been analysed
from a theoretical point of view.~\cite{BF06,BRF07} 
Adapting the standard effective-mass, (statically) screened 
Coulomb interaction model to the Tm[Se,Te] electron-hole system,  
the valence-band-hole conduction-band-electron mass asymmetry was found 
to suppress the excitonic insulator (EI) phase on the semimetallic side, 
as observed experimentally. But also on the semiconducting side,
the EI instability might be prevented---within this model---by either 
electron-hole liquid phases~\cite{BRF07,Ri77} or, at very large 
electron-hole mass ratios ($\gtrsim 100$), 
by Coulomb crystallization.~\cite{BFFLF05}      
The effective-mass Mott-Wannier-type exciton model neglects, however, 
important band structure effects, intervalley-scattering of
excitons, as well as exciton-phonon scattering . Moreover, the excitons 
in Tm[Se,Te] are rather small-to-intermediate sized bound objects  
(otherwise the experimentally estimated exciton density 
of about $1.3\times 10^{21}$ cm$^{-3}$ would lead to a 
strong overlap of the exciton 
wave functions, see Refs.~\onlinecite{NW90} and~\onlinecite{WBM04}). 
Hence the usual Mott-Wannier exciton description seems to be inadequate.

The onset of an EI phase was invoked quite recently 
in the transition-metal dichalcogenide $1T$-$\rm TiSe_2$
as driving force for the charge-density-wave (CDW) transition.~\cite{Ceea07} 

The perhaps minimal lattice model capable of describing the
generic two-band situation in materials being possible candidates 
for an EI scenario might be the Falicov-Kimball model (FKM), 
introduced about 40 years ago in order
to explain  the metal-insulator transition in certain transition-metal and 
rare-earth oxides.~\cite{FK69,RFK70} In its original form the model 
introduces two types of fermions: itinerant c (or d) 
electrons and localized f electrons with orbital energies $\varepsilon_c$ and
$\varepsilon_f$, respectively. The on-site Coulomb interaction 
between c- and f-electrons determines the distribution of 
electrons between these ``sub-systems'', and therefore may 
drive a valence transition as observed, e.g., in heavy-fermion 
compounds. To be a good model of the mixed-valence state, however, 
one should build in a coherence between c and f 
particles.~\cite{POS96a} This can be achieved by a c-f 
hybridization term $V$. Alternatively, a finite f-bandwidth,
which is certainly more realistic than entirely localized f-electrons, 
can also induce c-f coherence. 
The FKM with direct f-f hopping is sometimes called extended 
Falicov-Kimball model (EFKM). 

Most notably, it has been suggested that a novel ferroelectric 
state could be present in the mixed-valence phase of the FKM  
with hybridization.~\cite{POS96a} The origin is a non-vanishing 
$\langle c^\dagger f \rangle $ expectation value, causing a finite 
electrical polarization. In the limit $V\to 0$ there is no 
ferroelectric ground state, as was shown in Refs.~\onlinecite{Cz99}
and~\onlinecite{Fa99} in contrast to the findings in 
Ref.~\onlinecite{POS96a}. Afterwards it 
has been demonstrated that spontaneous electronic ferroelectricity 
also exists in the EFKM, provided that the c- and f-bands involved have 
different parity.~\cite{Ba02b}

By means of constrained path Monte Carlo 
techniques the ($T=0$) quantum phase diagram of the EFKM was calculated
in the intermediate-coupling regime for one- and two-dimensional (2D)  
systems, confirming the existence of a ferroelectric phase.~\cite{BGBL04} 
A more recent 2D Hartree-Fock  phase diagram of the EFKM~\cite{Fa08} 
was found to agree surprisingly well with the Monte Carlo data, 
supporting mean-field approaches to the 3D EFKM~\cite{Fa08,SC08} 
(on  the other hand the Hartree-Fock results have been 
questioned by a slave-boson treatment~\cite{Br08}). 
The ferroelectric state of the EFKM
can be viewed as an excitonic condensate ($\langle c^\dagger f \rangle $ 
is an excitonic expectation value since $f$ creates a f-band hole, 
i.e., the phase with non-vanishing polarization is in fact an EI phase).

Therefore, in this paper, we study the formation of excitonic bound states
and the nature of the EI phase in the framework of the (spinless) 3D EFKM.  
It can be written as 
\begin{equation}
H = \sum_{k\sigma} \varepsilon_{k\sigma} n_{k\sigma} +
U \sum_i n_{i\uparrow}n_{i\downarrow}\,,
\label{fkm}
\end{equation}
where f- and c-orbitals are labeled by the pseudospin variable
$\sigma=\uparrow,\downarrow$ (or $\sigma=\pm$) with 
$n_{k\sigma}^{}=a_{k\sigma}^{\dagger}a_{k\sigma}^{}$,
$a_{k\uparrow}\equiv f_k$, and $a_{k\downarrow}\equiv c_k$.
In Eq.~(\ref{fkm}), 
$\varepsilon_{k\sigma}=\varepsilon_{\sigma}+t_\sigma\gamma_k-\mu$,
$\gamma_k=\tfrac{1}{3}(\cos k_x+\cos k_y+ \cos k_z)$, and $\mu$ is the 
chemical potential. The signs of the transfer integrals $t_\sigma$ 
determine the type of the gap for large enough
$|\varepsilon_\uparrow-\varepsilon_\downarrow|$ and/or $U$ and of the electronic
insulator (ferroelectric) state. Provided that $t_\uparrow t_\downarrow < 0$ 
[$t_\uparrow t_\downarrow > 0$] we have a
direct [indirect] gap and the possibility of ferroelectricity (FE)
[antiferroelectricity (AFE)] with ordering vector $Q=0$ 
[$Q=(\pi,\pi,\pi)$]. 
In what follows, we put $\varepsilon_\uparrow =0$, 
$\varepsilon_\downarrow>0$, $t_\downarrow<0$, $t_\uparrow>0$, 
and consider the half-filled band case $\sum_\sigma n_\sigma =1$,
where $n_\sigma = \langle n_{i\sigma}\rangle = \tfrac{1}{N}\sum_k\langle 
n_{k\sigma}\rangle $.

First we investigate the existence of excitonic bound states in the 
phase without long-range order. To this end, we define the creation 
operator of a Frenkel-type exciton by
\begin{equation}
b_i^\dagger=a^\dagger_{i\downarrow}a^{}_{i\uparrow}\,,\quad
b_q^\dagger=\frac{1}{\sqrt{N}}\sum_k a^\dagger_{k+q\downarrow}
a^{}_{k\uparrow}\,,
\label{xop}
\end{equation} 
and the exciton commutator Green function
\begin{equation} 
G_X(q,\omega)=\langle\langle b_q^{};b_q^\dagger\rangle\rangle_\omega\,.
\label{gfb}
\end{equation}  
In the fermion representation of spins, we have 
$b_i^\dagger=S_i^-$, $b_q^\dagger=S_{-q}^-=(S^+_q)^+$,
so that $G_X(q,\omega)=\langle\langle S_q^{+};S_{-q}^-\rangle\rangle_\omega$
may be considered as the negative dynamic pseudospin susceptibility.
Therefore, to obtain the exciton propagator $G_X$, the calculation of the
dynamic spin susceptibility in the Hubbard model~\cite{GHE01} can be adopted.
Taking the random phase approximation we get
\begin{equation}
G_X(q,\omega)=\frac{G^{(0)}_X(q,\omega)}{1+UG^{(0)}_X(q,\omega)}\,,
\label{gx}
\end{equation}
where
\begin{equation}
G^{(0)}_X(q,\omega)=\frac{1}{N} \sum_k \frac{f(\bar{\varepsilon}_{k\uparrow})
-f(\bar{\varepsilon}_{k+q\downarrow})}{\omega-\omega_k(q)} \,,
\label{gxh}
\end{equation}
and
$\omega_k(q)=\bar{\varepsilon}_{k+q\downarrow}-\bar{\varepsilon}_{k\uparrow}$
describes the continuum of electron-hole excitations. 
Here, $\bar{\varepsilon}_{k\sigma}=\varepsilon_{k\sigma}+Un_{-\sigma}$ with
$\langle n_{k\sigma} \rangle =f(\bar{\varepsilon}_{k\sigma})$ and  
$f(\varepsilon)=[{\rm e}^{\varepsilon/T} +1]^{-1}$. Note that $G^{(0)}_X$
may be also obtained by the use of the Hamiltonian $H^{(0)}=\sum_{k\sigma}
\bar{\varepsilon}_{k\sigma} n_{k\sigma} $ resulting from the Hartree
decoupling of the interaction term in Eq.~(\ref{fkm}).

\begin{figure}[t]
  \centering 
  \includegraphics[width=\linewidth,clip]{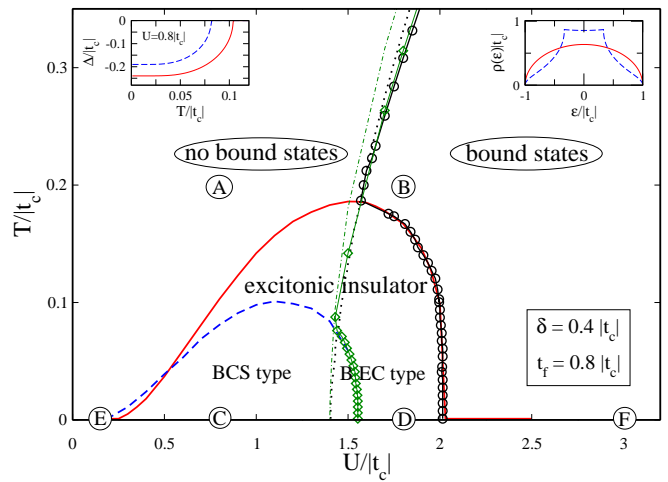}
  \caption{(color online) Phase boundary of the EI phase in the EFKM 
which typifies either a BEC or BCS condensate. Red solid (blue dashed) 
lines are obtained using the semielliptic (3D simple cubic) DOS shown in 
the right inset. The symbols give the critical $U$-values for exciton 
formation. Note that  preformed pairs may exist in the normal phase 
of the EFKM. Black dotted (green dashed-dotted) lines mark 
the opening of the gap $E_g^{(0)}$ (assuming $\Delta=0$). 
For further explanation see text. The left inset
shows exemplarily the suppression of the order parameter $\Delta$ 
with increasing temperature on the track from  C $\to$ A, i.e. at
fixed $U$, where a second-order phase transition is obtained.}
  \label{fig_pd}
\end{figure}

The exciton binding energy is obtained from the poles of $G_X(q,\omega)$ 
outside the continuum, i.e. from 
\begin{equation} 
-G^{(0)}_X(q,\omega)=U^{-1}
\label{xbe}
\end{equation}
with $0<\omega<\omega_k(q)|_{min}$. Let us consider excitons with $q=0$.
Then, we look for poles with $\omega\equiv \omega_X^{}<\omega_k(0)|_{min}=
E_g^{(0)}$,
where the gap $E_g^{(0)}$ is given by
 \begin{equation} 
E_g^{(0)}=\varepsilon_\downarrow
-|t_\downarrow|-t_\uparrow+U(n_\uparrow-n_\downarrow)\,.
\label{hgap}
\end{equation}
Here, we are mainly interested in the critical Coulomb attraction
$U_X(T)$ for the formation of excitonic bound states. 
Eq.~(\ref{xbe}) is solved numerically for $q=0$,
using both the semielliptic model density of states (DOS)  
$\rho(\varepsilon)=\tfrac{2}{\pi|t_\downarrow|}
[1-(\varepsilon/|t_\downarrow|)^2]^{1/2}$ (
$2|t_\downarrow|$  gives the bandwidth $W_\downarrow$)
and the tight-binding DOS for the 3D cubic lattice
(see right inset of Fig.~\ref{fig_pd}).

In Fig.~\ref{fig_pd} the boundary $U_X(T)$ for exciton
formation at $U>U_X(T)$ for both DOS models is plotted 
(circles, diamonds).
At low temperatures we find the boundary to be rather
sensitive to the shape of the DOS, whereas at higher temperatures
both $U_X(T)$ curves merge. Considering a fixed value
of $U>U_X(T=0)$, with increasing temperature the excitons
gradually dissociate into single holes and electrons, where
at $T>T_X(U)$ the bound states are lost. From the cusp  
of the boundary at the point ($T_s, U_s$) we may suggest an instability against
a homogeneous phase with long-range order at $T<T_s$.

The $T=0$ Hartree-Fock ground-state phase diagram of the 3D 
half-filled EFKM~\cite{Fa08,SC08} exhibits---besides full 
f-band and c-band insulator 
regions at large splittings $\delta=|\varepsilon_c-\varepsilon_f|$---two 
symmetry-broken states: the anticipated excitonic insulator and a 
CDW. While the CDW ground state is stable for all 
ratios $t_f/t_c$ at $\delta=0$, it becomes rapidly suppressed 
for $\delta >0$, especially if the c- and f-bandwidths 
are comparable.~\cite{Fa08} 
Since we are interested in the (homogeneous) condensed excitonic 
phase only, we adjust the parameters $\delta$, $t_f$, and $U$ accordingly. 
To model the intermediate-valence situation we choose 
$\delta/|t_c|=0.4$ and $t_f/|t_c|=0.8$.  
The almost perfect agreement between the (2D) Hartree-Fock and
path Monte Carlo phase diagrams for intermediate couplings~\cite{Fa08} 
might justify the application of the Hartree-Fock 
approach to values of the Coulomb attraction $U$ of the order 
of the bandwidth. 

To make contact with previous Hartree-Fock 
approaches~\cite{Fa08,SC08}, we use the equation of 
motion method for the anticommutator Green functions~\cite{GHE01} 
$\langle\langle a^{}_{k\sigma};a^\dagger_{k\sigma}\rangle\rangle_\omega$
and $\langle\langle a^{}_{k\sigma};a^\dagger_{k,-\sigma}\rangle\rangle_\omega$,
and perform a decoupling that allows for the description
of the FE EI phase by the order parameter
\begin{equation}
\Delta=\frac{U}{N}\sum_k\langle a^\dagger_{k\uparrow}a^{}_{k\downarrow}\rangle\,.
\label{op}
\end{equation}
We obtain 
$\langle a^{\dagger}_{k\uparrow} a^{}_{k\downarrow} \rangle
= \Delta \sum_\sigma \sigma f(E_{k\sigma})/(2E_k)$
with $E_{k\sigma}=\tfrac{1}{2}(\bar{\varepsilon}_{k\downarrow} 
+\bar{\varepsilon}_{k\uparrow})-\sigma E_k$,
$E_k=[\xi_k^2+\Delta^2]^{1/2}$,
$\xi_k=\tfrac{1}{2}(\bar{\varepsilon}_{k\downarrow} 
-\bar{\varepsilon}_{k\uparrow})$,  and
\begin{equation}
\langle n^{}_{k\sigma}\rangle=
\frac{1}{2}\left[1+\sigma\frac{\xi_k}{E_k}\right] f(E_{k\uparrow}) 
+ \frac{1}{2}\left[1-\sigma\frac{\xi_k}{E_k}\right] f(E_{k\downarrow})\,. 
\label{nks}
\end{equation}
Then, for $\Delta\neq 0$, we get the gap equation 
\begin{equation}
1 = \frac{U}{N}   \sum_{k\sigma} \sigma \frac{f(E_{k\sigma})}{2E_k}\,. 
\label{gap}
\end{equation}

Figure~\ref{fig_pd} shows the finite-temperature phase boundary 
of the EI phase obtained by the self-consistent 
solution of the Hartree-Fock Eqs.~(\ref{op})--(\ref{gap}). 
In comparison with the semielliptic DOS (solid line), the use of the more 
realistic tight-binding DOS (dashed line) yields a shrinking of 
the EI phase, which corresponds to the behavior of the boundary 
$U_X(T)$ for  exciton formation. 
For $U>U_s$ we obtain the phase boundary $T_c(U)$ coinciding
with the boundary $T_X(U)$ for exciton formation. This result
gives a strong argument for the BEC of preformed (tightly bound)
excitons at $T_c(U>U_s)$. On the other hand, for $U<U_s$ there are no 
preformed excitons above $T_c$, and a BCS-like condensation  
at $T_c(U<U_s)$ takes place, i.e., the pair formation and condensation 
occurs simultaneously. 
Although the gap equation captures the BCS and BEC situation 
at weak and strong couplings~\cite{NS85,BF06}, it cannot
discriminate between them.  

Thus, the existence or non-existence of bound states 
above $T_c$ gives strong evidence for a BEC or BCS transition 
scenario at $T_c$, respectively. Moreover, within the EI phase, 
a crossover from a strong-coupling BEC to a weak-coupling BCS 
condensate of electron-hole pairs is strongly suggested.

To describe qualitatively this BEC-BCS crossover region, we consider the gap 
boundary $U_g(T)$ (thin dotted and dashed-dotted lines) resulting from
$E_g^{(0)}=0$ [Eq.~(\ref{hgap})], where the gap opens for $U>U_g$.
Interestingly, at $T_s=T_c(U_s)$ we obtain $U_g(T_s)=U_X(T_s)$.
That is, at this point the opening of the gap is accompanied with 
the formation of bound states, whereas
for $T>T_c$, $U_g(T)$ is slightly smaller than $U_X(T)$. From this
result we may get a crude estimate of the BEC-BCS crossover 
region by extrapolating $U_g(T)$ into the EI phase. 
Solving Eq.~(\ref{xbe}) at a fixed $T < T_c$, for $U$ in 
the region $U_g < U < U_X$ we get negative pole energies $\omega_X$ 
which indicates the instability of the normal phase with bound states 
against the long-range ordered EI phase. Moreover, for 
$U\lesssim U_g$ no solution can be found which may be indicative 
for an instability towards a BCS-type EI state. Thus, 
the BEC-BCS crossover in the EI phase should occur in the 
neighborhood of the $U_g(T)$ line.

In comparison to the phase boundary obtained 
within the simple effective-mass, Mott-Wannier-type model~\cite{BF06,BRF07} 
the EI phase of the EFKM  is confined  
at zero temperature on the weak-coupling side, because 
of the finite  f- and c-bandwidths. While the shape of the EI dome 
approximates the Tm[Te,Se] phase diagram constructed from 
the experimental data, the absolute transition temperatures are 
overestimated, of course, by any mean-field approach. The homogeneous
EI phase shrinks as $\delta$ as $t_f$ becomes smaller 
at fixed $\delta$, but it does not disappear.~\cite{Fa08}

\begin{figure}[t]
  \centering 
  \includegraphics[width=\linewidth,clip]{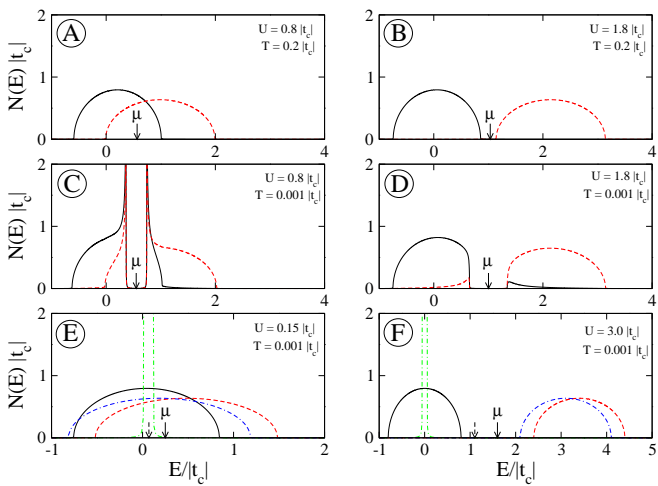}
  \caption{(color online) DOS for f-band ([black] 
solid curves) and c-band electrons ([red] dashed curves) at the  
points marked by A to F in  the $T$-$U$ 
plane in Fig.~\ref{fig_pd}. Band structure 
parameters are  $\delta=0.4 |t_c|$, $t_f=0.8 |t_c|$.
For comparison, the results for $\delta=0.1 |t_c|$ and $t_f=0.05 |t_c|$ 
([green] double-dot-dashed and [blue] dot-dashed curves) 
have been included in panels E and F; here a CDW
will become the Hartree-Fock ground state as $U$ increases from E to F.  
All data are obtained for the semielliptic DOS and total 
filling $n=n_f+n_c=1$; the arrows mark the positions 
of the chemical potential. }
  \label{fig_dos}
\end{figure}

Figure~\ref{fig_dos} gives the partial f- and c-electron DOS 
at various characteristic points A-F of the phase diagram 
shown in Fig.~\ref{fig_pd}. The high-temperature phase may be
viewed as a metal/semimetal (panel A) 
or a semiconductor (panel B) in the weak- or intermediate-to-strong 
interaction regime, respectively. The EI phase shows completely different
behavior. As can be seen from panel C, a correlation-induced 
``hybridization'' gap opens, indicating long-range order 
(non-vanishing f-c-polarization). As the temperature increases 
the gap weakens and finally closes at $T=T_c$.
The pronounced c-f-state mixing and strong enhancement of the 
DOS at the upper/lower valence/conducting band edges
is reminiscent of a BCS-like structure evolving from a 
(semi-) metallic state with a large Fermi surface above $T_c$
(see panel A). This may be in favor of a BCS pairing in the weak-coupling
region of the EI phase, as discussed above. By contrast the DOS 
shown in panel~D clearly evolves from a gapped high-temperature phase.
Finally, in panel E [F] the partial f- and c-electron DOS at $T\simeq 0$
below [above] the EI phase are depicted, where the system behaves 
as a metal or semimetal [band insulator or semiconductor].
Note that the splitting of c- and f-bands in panel F is not caused 
by $\delta$ (being the same as in E), but is due to the Hartree 
shift $\propto U (n_f-n_c)$. 

To summarize, in this work, we attempted to link experimental
hints for excitonic condensation  to recent theoretical studies of electronic 
ferroelectricity in the extended Falicov-Kimball model. We analyzed
the finite-temperature phase diagram and argued that a finite f-bandwidth
in combination with a short-range interband Coulomb attraction between
(heavy) valence-band holes and (light) conduction-band electrons may 
lead to f-c-band coherence and an excitonic insulator low-temperature phase.  
Most noteworthy we established the existence of excitonic bound states
for the EFKM on the semiconductor side of the 
semimetal semiconductor transition 
above $T_c$, and suggested a BCS-BEC crossover scenario within the 
condensed state. As a consequence, we expect pronounced transport
anomalies in the transition regime both in the low- and 
high-temperature phases, which should be studied in the framework 
of the EFKM in future work.

{\it Acknowledgments.}
This work was supported by DFG through SFB 652. 
The authors thank B. Bucher and A. Wei{\ss}e for helpful discussions.

\end{document}